\documentstyle[epsf,twocolumn,fleqn]{article}

\newcommand{\CGO}{$\rm CuGeO_3$}

\begin{document}
\renewcommand{\textfraction}{0.1}
\renewcommand{\topfraction}{0.8}
\rule[-8mm]{0mm}{8mm}
\begin{minipage}[t]{15cm}
\begin{center}
{\LARGE \bf Lattice Dimerization in the Spin-Peierls 
Compound CuGeO$_{\bf 3}$\\[4mm]}
B.~B\"uchner$^{\rm a}$, H.~Fehske$^{\rm b}$, 
A.P.~Kampf$^{\rm\, c}$, and G.~Wellein$^{\rm b}$\\[3mm]
$^{\rm a}${II. Physikalisches Institut, Universit\"at zu K\"oln, 
50937 K\"oln, Germany}\\
$^{\rm b}${Physikalisches Institut, Universt\"at Bayreuth, 
95440 Bayreuth, Germany}\\
$^{\rm c}${Institut f\"ur Physik, Universit\"at Augsburg, 86135 Augsburg, 
Germany}\\[4.5mm]
\end{center}
{\bf Abstract}\\[0.2cm]
\hspace*{0.5cm}
The uniaxial pressure dependences of the exchange coupling and 
the structural distortion in the dimerized phase of  CuGeO$_{\bf 3}$
are analyzed. A minimum magnetic dimerization of 3 \% is obtained, 
incompatible with an adiabatic approach to the spin--Peierls transition. 
Exploring the properties of an Heisenberg spin chain with
{\it dynamical} spin-phonon coupling, the dimerization dependence of the 
spin excitation gap is found to be in qualitative agreement with 
experiment. \\[0.2cm]
\end{minipage}\\[4.5mm]
\normalsize
There has been a renewed interest in the spin--Peierls (SP) 
phenomenon since it was recognized that the displacive SP 
transition in the first inorganic SP compound {\CGO} 
shows no phonon softening. On the contrary, the Peierls--active
optical phonon modes with frequencies $\omega\approx J$ and 
$\omega\approx 2J$ ($J$ is the exchange coupling 
between nearest neighbour (nn) Cu spins) 
harden by about 5 \% with decreasing 
temperature, requiring a very strong spin-phonon coupling for the 
SP instability to occur~\cite{BradenPRB}.
  
In spite of this fact, up to now most theoretical studies rely 
on an adiabatic approach to the phonons. The Hamiltonian 
commonly used to model the dimerized (D) SP phase is
\begin{equation}
{\cal H} = \sum_{j} J\left[(1 + (-1)^{j}\delta){\bf S}_j\cdot{\bf S}_{j+1}\right] +
J'{\bf S}_j\cdot{\bf S}_{j+2},
\label{hamil}
\end{equation}
where $\delta$ describes the alternating pattern of weak and strong bonds
due to the structural distortion and $J'$ is the next nn coupling.
From the magnetic properties of the uniform (U) phase $J\simeq 160$~K and 
$\alpha=J'/J\simeq 0.35$ have been estimated for {\CGO}~\cite{fabricius}. 
However, if one attempts to reproduce the observed spin gap 
$\Delta^{ST}\simeq 2.1~\rm{meV}$ from~(\ref{hamil}), 
a very small value of $\delta\simeq 0.012$ results.
 
An alternative and model independent way to determine $\delta$ is possible by 
combining the amount of the structural distortion in the D-phase with the 
magneto-elastic coupling in the U-phase using the uniaxial pressure derivatives 
of the magnetic susceptibility at high temperatures as measured by 
magnetostriction~\cite{bbprl}. Rather large values $\partial\ln J/\partial p_i$
of 4, -9, and -2 \%/GPa are obtained for pressures $p_i$ along the $a$, $b$, 
and $c$ axis, respectively, whereas $\partial J'/\partial p$ is essentially 
zero~\cite{fabricius}.
\begin{figure}[t]
\unitlength1mm
\begin{picture}(70,80)
\end{picture}
\end{figure}
We stress that these results do not depend crucially on the values of $J$ and
$J'$. The decrease of $J$ for pressure applied parallel to the spin chains 
($c$ axis) signals the correlation between $J$ and the Cu-O-Cu bond angle 
$\gamma$~\cite{BradenPRB,bbprl,GeeKhom}.

The $c$-axis lattice constant depends on $\gamma$ only and the Cu-O bond 
length $d_{CO}$ ($c=2d_{CO}\sin(\gamma/2)$). Obviously 
$\partial d_{CO}/\partial p_c\le 0$, $\partial J/\partial d_{CO}\le 0$, and 
from Ref.~\cite{BBxx} we have $\partial\ln c/\partial p_c\simeq -0.3$~\%/GPa. 
The observed negative 
$\partial J/\partial p_c$ is due to a positive $\partial J/\partial\gamma$
which is the most important source of spin-phonon coupling in 
{\CGO}~\cite{BradenPRB}. By accounting for the 
bond angle and length dependences 
a clear-cut lower limit $\partial\ln J/\partial\gamma\le 5\%/^\circ$ is 
obtained for the angular dependence of $J$. The true value is, however, much 
larger (of order $10\%$) for two reasons: Uniaxial pressure also reduces
$d_{CO}$ causing an additional positive pressure effect on $J$ and the elastic 
reactions of the two other lattice constants also increase 
$\partial\ln J/\partial\gamma$.

Qualitatively the positive $\partial J/\partial p_a$ follows from
$\partial\ln d_{CO}/\partial p_a\le 0$ and $\partial\gamma/\partial p_a\ge 0$.
In order to explain the very large negative pressure derivative with respect to
$p_b$ we have to take into account the hybridization between Ge and O 
orbitals~\cite{GeeKhom}. This side group effect depends on the angle 
$\phi\simeq 159^\circ$ between the Ge-O bonds and the $\rm CuO_4$ 
plaquettes~\cite{GeeKhom}. The large compressibility of the $b$ axis is related
to a decrease of $\phi$~\cite{BradenPRB,BBxx} and explains the strong reduction
of $J$ for $p\|b$; we estimate $\partial\ln J/\partial\phi\simeq 1~\%/^\circ$.

In the D-phase of {\CGO} the structural dimerization is related to two
independent phonon modes~\cite{BradenPRB}. Both, $\gamma$ and
$\phi$, alternate in the dimerized phase ($\gamma_1-\gamma_2=0.83^\circ$ and 
$\phi_1-\phi_2=1.8^\circ$), whereas $d_{CO}$ hardly differs for neighboring 
exchange paths~\cite{BradenPRB}. From the lower bounds of the coupling 
constants a minimum magnetic dimerization of $(J_1-J_2)/2J\equiv\delta\ge 3\%$
is obtained. Thus the small value $\delta=1.2\%$, which is 
necessary to reproduce the correct spin gap within the static model 
Eq.~(\ref{hamil}), is clearly excluded from our data analysis. 

The simplest model that includes lattice dynamical effects may be obtained
from~(1) by replacing~\cite{WFK98,APSA98} 
\begin{equation}
(-1)^i\delta \to g u_i = g (b_i^{\dagger} + b_i^{})\,, 
\end{equation}
where $g$ is the magneto-elastic coupling constant and the 
$b_i^{(\dagger)}$ are independent phonon 
destruction (creation) operators on each bond $(i,i+1)$. 
For modeling the optical phonons in {\CGO}
we choose dispersionless Einstein modes with frequency $\omega$, 
i.e. ${\cal H}_{ph}= \omega \sum_i b_i^{\dagger} b_i^{}$.  
Recently it was shown that such a dynamical spin-phonon model 
describes the general features of the magnetic excitation spectrum
of {\CGO}~\cite{WFK98}. 

Here we focus on the behavior of the lattice dimerization which can
be found from the displacement structure factor at $q=\pi$: 
\begin{equation}
\delta^2 = \frac{g^2}{N^2}\sum_{i,j} \langle u_i u_j \rangle 
\mbox{e}^{i\pi(R_i-R_j)}\,.
\end{equation}
Since in the physically most relevant {\it non-adiabatic} 
($\omega\stackrel{>}{\sim} J$) and 
{\it intermediate coupling} ($g\sim J$) regime 
the spin and phonon dynamics are intimately related, 
we performed a complete numerical diagonalization of the 
quantum phonon model~(2) using a controlled
phonon Hilbert space truncation~\cite{WFK98}.
  
Fig.~1 summarizes our main results:   
(i) In the non-adiabatic region the SP transition takes place
provided that $(g/\omega) > (g/\omega)_c \simeq 1$, 
{\it irrespective} of the ratio $\omega/J$. 
For {\CGO} with $\omega\sim J$ this implies $g > g_c\simeq J$, i.e.
a strong-coupling situation. (ii) Below (above) $(g/\omega)_c$
the dimerization decreases (increases) with increasing lattice size
(see lower inset), indicating that the {\it infinite} system~(2) exhibits 
a true phase transition. (iii) The dynamic model~(2) partially resolves
the $\Delta^{ST}-\delta$ conflict we are faced within the static approach~(1),
because the dimerization $\delta$ {\it grows} with the phonon frequency  
at fixed $\Delta^{ST}$ (cf. upper inset). Thus matching $\Delta^{ST}$ 
to the neutron scattering data, a larger $\delta\simeq$ 5\% may result. 

In conclusion, our findings are in qualitative agreement with 
experimental data on {\CGO}, suggesting the necessity 
for a non-adiabatic spin-phonon approach to this material. 
\begin{figure}[t]
\vspace*{-0.5cm}
\centerline{\mbox{\epsfxsize 8.5cm\epsffile{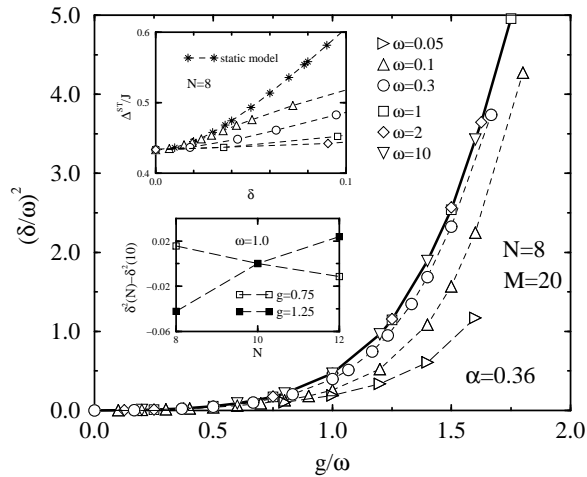}}}\vspace*{-0.5cm}
\caption{\small Dimerization in the frustrated Heisenberg model 
with dynamic spin-phonon coupling. Results are presented for an 
eight-site chain with M=20 phonons at different 
phonon frequencies.}\vspace*{-0.3cm}
\end{figure} 
\vspace*{-0.4cm}
{\small
\begin{thebibliography}{99}\vspace*{-0.2cm}
\bibitem{BradenPRB} 
M. Braden et al., Phys. Rev. B{\bf 57}, 1105 (1996); 
Phys. Rev. Lett. {\bf 80}, 3634 (1998).\vspace*{-0.1cm}
\bibitem{fabricius}
K. Fabricius et al., Phys. Rev. B{\bf 57}, 1102 (1997).\vspace*{-0.1cm}
\bibitem{bbprl}
U. Ammerahl et al., Z. Phys. B{\bf 102}, 71 (1997).\vspace*{-0.1cm}
\bibitem{GeeKhom} 
W. Geertsma$\!$ et$\!$ al., Phys.$\!$ Rev.$\!$ B{\bf 54}, 3011 (1996).\vspace*{-0.1cm}
\bibitem{BBxx} 
M. Saint-Paul et al., Phys. Rev. B{\bf 52}, 15298 (1995); S. Br\"auninger et 
al., Phys. Rev. B{\bf 56}, 11357 (1997).\vspace*{-0.1cm}
\bibitem{WFK98}
G. Wellein et al., cond-mat/9804085.\vspace*{-0.1cm}
\bibitem{APSA98}
D. Augier et al., cond-mat/9802053.
\end{thebibliography}
}
\end{document}